\documentclass[10pt,letterpaper,final]{iopart}

\usepackage{bbm,amssymb,amsbsy,graphicx,subfigure,times}

\usepackage[latin1]{inputenc}
\newcommand{\R}{\mathbbm{R}}

\newcommand{\rr}{\mathbbm{R}}

\newcommand{\id}{\mathbbm{1}}

\newcommand{\tn}{\tilde{\nu}_-}
\newcommand{\sy}[1]{Sp_{(#1,\R)}}

\renewcommand{\tr}{{\rm Tr}\,}
\renewcommand{\det}{{\rm Det}\,}
\newcommand{\gr}[1]{\boldsymbol{#1}}
\newcommand{\be}{\begin{equation}}
\newcommand{\ee}{\end{equation}}
\newcommand{\bea}{\begin{eqnarray}}
\newcommand{\eea}{\end{eqnarray}}
\newcommand{\ket}[1]{|#1\rangle}
\newcommand{\bra}[1]{\langle#1|}

\newcommand{\N}{{\cal N}}

\newcommand{\sig}{\gr{\sigma}}
\newcommand{\gam}{\gr{\gamma}}
\newcommand{\eps}{\gr{\varepsilon}}

\newcommand{\eq}[1]{Eq.~(\ref{#1})}

\newcommand{\ie}{\emph{i.e.}~}

\begin{document}
\title{Optical implementation and entanglement distribution in Gaussian valence bond states}
\date{February 6, 2007}
\author{Gerardo Adesso}
\address{Centre for Quantum Computation, DAMTP, Centre for
Mathematical Sciences, University of Cambridge, Wilberforce Road,
Cambridge CB3 0WA, United Kingdom; \\
Dipartimento di Fisica, Universit\`a di Roma ``La
Sapienza'', Piazzale Aldo Moro 5, 00185 Roma, Italy; \\
Dipartimento di Fisica ``E. R. Caianiello'', Universit\`a degli Studi di Salerno, Via S. Allende, 84081
Baronissi (SA), Italy}
\author{Marie Ericsson}
\address{Centre for Quantum Computation, DAMTP, Centre for
Mathematical Sciences, University of Cambridge, Wilberforce Road,
Cambridge CB3 0WA, United Kingdom}


\begin{abstract}
We study Gaussian valence bond states of continuous variable
systems, obtained as the outputs of projection operations from an
ancillary space of $M$ infinitely entangled bonds connecting
neighboring sites, applied at each of $N$ sites of an harmonic
chain. The entanglement distribution in Gaussian valence bond states
can be controlled by varying the input amount of entanglement
engineered in a $(2M+1)$-mode Gaussian state known as the building
block, which is isomorphic to the projector applied at a given site.
We show how this mechanism can be interpreted in terms of multiple
entanglement swapping from the chain of ancillary bonds, through the
building blocks. We provide optical schemes to produce bisymmetric
three-mode Gaussian building blocks (which correspond to a single
bond, $M=1$), and study the entanglement structure in the output
Gaussian valence bond states. The usefulness of such states for
quantum communication protocols with continuous variables, like
telecloning and teleportation networks, is finally discussed.
\end{abstract}
\maketitle

\section{Introduction}

Quantum information aims at the treatment and transport of
information in ways forbidden by classical physics. For this goal,
continuous variables (CV) of atoms and light have emerged as a
powerful tool \cite{review}. In this context, entanglement is an
essential resource. Recently, the valence bond formalism, originally
developed for spin systems \cite{maria}, has been generalized to the
CV scenario \cite{GMPS,marie} for the special class of Gaussian
states, which play a central role in theoretical and practical CV
quantum information and communication \cite{adebook}.

In this work we analyze feasible implementations of Gaussian valence
bond states (GVBS) for quantum communication between many users in a
CV setting, as enabled by their peculiar structure of distributed
entanglement \cite{marie}. After recalling the necessary notation
(Sec.~\ref{seccv}) and the construction of Gaussian valence bond
states \cite{GMPS} (Sec.~\ref{secgvbs}), we discuss the
characterization of entanglement and its distribution in such states
as regulated by the entanglement properties of simpler states
involved in the valence bond construction \cite{marie}
(Sec.~\ref{secdist}). We then focus on the realization of GVBS by
means of quantum optics, provide a scheme for their state
engineering (Sec.~\ref{secopt}), and discuss the applications of
such resources in the context of CV telecloning
\cite{teleclon,teleclonexp} on multimode harmonic rings
(Sec.~\ref{secclon}).

\section{Continuous variable systems and Gaussian
states}\label{seccv}

A CV system \cite{review,adebook} is described by a Hilbert space
${\cal H}=\bigotimes_{i=1}^{N} {\cal H}_{i}$ resulting from the
tensor product of infinite dimensional Fock spaces ${\cal H}_{i}$'s.
Let $a_{i}$ and $a_i^\dag$ be the annihilation and creation
operators acting on ${\cal H}_{i}$ (ladder operators), and $\hat
q_{i}=(a_{i}+a^{\dag}_{i})$ and $\hat p_{i}=(a_{i}-a^{\dag}_{i})/i$
be the related quadrature phase operators.   Let $\hat R= (\hat
x_{1},\hat p_{1},\ldots,\hat q_{N},\hat p_{N})$ denote the vector of
the operators $\hat q_{i}$ and $\hat p_{i}$. The canonical
commutation relations for the $\hat R_{i}$ can be expressed in terms
of the symplectic form ${\Omega}$ as
\[
[\hat R_{i},\hat R_j]=2i\Omega_{ij} \; ,
\]
\[
{\rm with}\quad{\Omega}\equiv {\omega}^{\oplus N}\; , \quad
{\omega}\equiv \left( \begin{array}{cc}
0&1\\
-1&0
\end{array}\right) \; .
\]

The state of a CV system can be equivalently described by
quasi-probability distributions defined on the $2N$-dimensional
space associated to the quadratic form $\Omega$, known as quantum
{\em phase space}. In the phase space picture, the tensor product
${\cal H}=\bigotimes_i^N{\cal H}_{i}$ of the Hilbert spaces ${\cal
H}_{i}$'s of the $N$ modes results in the direct sum
$\Lambda=\bigoplus_i^N\Lambda_{i}$ of the phase spaces
$\Lambda_{i}$'s.

States with Gaussian quasi-probability distributions are referred to
as {\em Gaussian states}. Such states are at the heart of
information processing in CV systems \cite{review,adebook} and are
the subject of our analysis. By definition, a Gaussian state is
completely characterized by the first and second statistical moments
of the field operators, which will be denoted, respectively, by the
vector of first moments $\bar R\equiv\left(\langle\hat R_{1}
\rangle,\langle\hat R_{2}\rangle,\ldots,\langle\hat R_{2N-1}\rangle,
\langle\hat R_{2N}\rangle\right)$ and the covariance matrix (CM)
$\gam$ of elements
\begin{equation}
\gamma_{ij}\equiv\frac{1}{2}\langle \hat{R}_i \hat{R}_j + \hat{R}_j
\hat{R}_i \rangle - \langle \hat{R}_i \rangle \langle \hat{R}_j
\rangle \, . \label{covariance}
\end{equation}

Coherent states, resulting from the application of displacement
operators $D_Y = e^{ i Y^T \Omega \hat R}$ ($Y \in \rr^{2n}$) to the
vacuum state, are Gaussian states with CM $\gam=\id$ and first
statistical moments $\bar R = Y$. First moments can be arbitrarily
adjusted by local unitary operations (displacements), which cannot
affect any property related to entropy or entanglement. They can
thus be assumed zero without any loss of generality. A $N$-mode
Gaussian state will be completely characterized by its real,
symmetric, $2N \times 2N$ CM $\gam$.

The canonical commutation relations and the positivity of the
density matrix $\rho$ of a Gaussian state imply the {\em bona fide}
condition
\begin{equation}
\gam+ i\Omega\ge 0 \; , \label{bonfide}
\end{equation} as a necessary and sufficient constraint the
matrix $\gam$ has to fulfill to be a CM corresponding to a physical
state \cite{simon87,simon}. Note that the previous condition is
necessary for the CM of {\em any} (generally non Gaussian) state, as
it generalizes to many modes the Robertson-Schr\"odinger uncertainty
relation \cite{seraliano}.

A major role in the theoretical and experimental manipulation of
Gaussian states is played by unitary operations which preserve the
Gaussian character of the states on which they act. Such operations
are all those generated by terms of the first and second order in
the field operators.  As a consequence of the Stone-Von Neumann
theorem, any such operation at the Hilbert space level corresponds,
in phase space, to a symplectic transformation, i.e.~to a linear
transformation $S$ which preserves the symplectic form $\Omega$, so
that $\Omega=S^T \Omega S$, i.e.~it preserves the commutators
between the different operators. Symplectic transformations on a
$2N$-dimensional phase space form the (real) symplectic group,
denoted by $\sy{2N}$. Such transformations act linearly on first
moments and ``by congruence'' on the CM ({\ie}so that $\gam\mapsto S
\gam S^T$). One has $\det{S}=1$, $\forall\,S\in\sy{2N}$. A crucial
symplectic operation is the one achieving the normal mode
decomposition. Due to Williamson theorem \cite{williamson36}, any
$N$-mode Gaussian state can be symplectically diagonalized in phase
space, so that its CM is brought in the form
 $\gr\nu$, such that $S \gr{\gamma} S^T=\gr{\nu}$,
with $\gr{\nu}=\,{\rm diag}\,\{\nu_1,\nu_1,\ldots\nu_N,\nu_N\}$. The
set $\{\nu_i\}$ of the positive-defined eigenvalues of
$|i\Omega\gr{\gamma}|$ constitutes the symplectic spectrum of
$\gr{\gamma}$ and its elements, the so-called symplectic
eigenvalues, must fulfill the conditions $\nu_i\ge 1$, following
from the uncertainty principle \eq{bonfide} and ensuring positivity
of the density matrix $\rho$ corresponding to $\gr{\gamma}$.

Ideal beam splitters, phase shifters and squeezers are described by
symplectic transformations. In particular, a phase-free two-mode
squeezing transformation, which corresponds to squeezing the first
mode (say $i$) in one quadrature (say momentum, $\hat p_i$) and the
second mode (say $j$) in the orthogonal quadrature (say position,
$\hat q_j$) with the same degree of squeezing $r$, can be
represented in phase space by the symplectic transformation
\begin{equation}\label{squeezing}
S_{ij}(r)={\rm diag}\{\exp{r},\,\exp{-r},\,\exp{-r},\,\exp{r}\}\,.
\end{equation}
These trasformations occur for instance in parametric down
conversions \cite{francussi}.  Another important example of
symplectic operation is the ideal (phase-free) beam splitter, which
acts on a pair of modes $i$ and $j$ as \cite{network}
$$
\hat{B}_{ij}(\theta):\left\{
\begin{array}{l}
\hat a_i \mapsto \hat a_i \cos\theta + \hat a_j\sin\theta \\
\hat a_j \mapsto \hat a_i \sin\theta - \hat a_j\cos\theta \\
\end{array} \right.
$$
and corresponds to a rotation in phase space of the form
\begin{equation}\label{bs}
B_{ij}(\theta)=\left(\begin{array}{cccc}
\cos(\theta)&0&\sin(\theta)&0\\
0&\cos(\theta)&0&\sin(\theta)\\
\sin(\theta)&0&-\cos(\theta)&0\\
0&\sin(\theta)&0&-\cos(\theta)
\end{array}\right)\,.
\end{equation}
The transmittivity $\tau$ of the beam splitter is given by
$\tau=\cos^2(\theta)$ so that a 50:50 beam splitter ($\tau=1/2$)
amounts to a phase-space rotation of $\pi/4$.

The combined application of a two-mode squeezing and a 50:50 beam
splitter
 realizes the entangling {\em twin-beam}
transformation \cite{wolfopt}
\begin{equation}\label{twin}
T_{ij}(r) =B_{ij}(\pi/4) \cdot S_{ij}(r)\,,
\end{equation}
which, if applied to two uncorrelated vacuum modes $i$ and $j$
(whose initial CM is the identity matrix), results in the production
of a pure two-mode squeezed Gaussian state with CM
$\gr\sigma_{i,j}(r) = T_{ij}(r) T_{ij}^T(r)$ given by
\begin{equation}\label{tmss}
\sig_{i,j}(r)=\left(\begin{array}{cccc}
\cosh(2r)&0&\sinh(2r)&0\\
0&\cosh(2r)&0&-\sinh(2r)\\
\sinh(2r)&0&\cosh(2r)&0\\
0&-\sinh(2r)&0&\cosh(2r)
\end{array}\right)\,.
\end{equation}

The CV {\em entanglement} in the state $\sig_{i,j}(r)$ increases
unboundedly as a function of $r$, and in the limit $r \rightarrow
\infty$ \eq{tmss} approaches the (unnormalizable)
Einstein-Podolski-Rosen (EPR) state \cite{epr}, simultaneous
eigenstate of relative position and total momentum of the two modes
$i$ and $j$. Concerning entanglement in general, the ``positivity of
partial transposition'' (PPT) criterion states that a Gaussian CM
$\gam$ is separable (with respect to a $1 \times N$ bipartition) if
and only if the partially transposed CM $\tilde{\gam}$  satisfies
the uncertainty principle \eq{bonfide} \cite{simon,werwolf}. In
phase space, partial transposition amounts to a mirror reflection of
one quadrature associated to the single-mode partition. If
$\{\tilde{\nu}_i\}$ is the symplectic spectrum of the partially
transposed CM $\tilde{\gr{\gamma}}$, then a $(N+1)$-mode Gaussian
state with CM $\gr{\gamma}$ is separable if and only if
$\tilde{\nu}_i\ge 1$ $\forall\, i$. A proper measure of CV
entanglement is the logarithmic negativity $E_{\N}$ \cite{vidwer},
which is readily computed in terms of the symplectic spectrum
$\tilde{\nu}_i$ of $\tilde{\gr{\gamma}}$ as
\begin{equation}\label{logneg}
E_{\N}=-\sum_{i:\,\tilde{\nu}_i<1}\log \tilde{\nu}_i\,.
\end{equation} Such an entanglement monotone \cite{plenio}
quantifies the extent to which the PPT condition $\tilde{\nu}_i\ge
1$ is violated. For $1 \times N$ Gaussian states, only the smallest
symplectic eigenvalue $\tn$ of the partially transposed CM can be
smaller than one \cite{seraliano}, thus simplifying the expression
of $E_{\N}$: then the PPT criterion simply yields that $\gr\gamma$
is entangled as soon as $\tilde{\nu}_{-}<1$, and infinite
entanglement (accompanied by infinite energy in the state) is
reached for $\tilde{\nu}_{-} \rightarrow 0^+$.

For $1 \times 1$ Gaussian states $\gr\gamma_{i,j}$ symmetric under
mode permutations, the entanglement of formation $E_F$ is computable
as well via the formula \cite{efprl}
\begin{equation}\label{eof}
E_F (\gr\gamma_{i,j}) = \max \{0,\, f(\tilde{\nu}^{i,j}_{-})\}\,,
\end{equation}
with $$f(x) = \frac{(1+x)^2}{4x} \log{\frac{(1+x)^2}{4x}} -
\frac{(1-x)^2}{4x} \log{\frac{(1-x)^2}{4x}}\,.$$ Being a
monotonically decreasing function of the smallest symplectic
eigenvalue $\tilde{\nu}^{i,j}_{-}$ of the partial transpose
${\tilde{\gr\gamma}}_{i,j}$ of $\gr\gamma_{i,j}$, the entanglement
of formation is completely equivalent to the logarithmic negativity
in this case. For a two-mode state, $\tilde\nu_{i,j}$ can be
computed from the symplectic invariants of the state
\cite{extremal}, and experimentally estimated with measures of
global and local purities \cite{apriori} (the purity $\mu=\tr\rho^2$
of a Gaussian state $\rho$ with CM $\gam$ is equal to $\mu =
(\det\gam)^{-1/2}$).

\section{Gaussian valence bond states}\label{secgvbs}

Let us review the basic definitions and notations for GVBS, as
adopted in Ref.~\cite{marie}. The so-called matrix product Gaussian
states introduced in Ref.~\cite{GMPS} are $N$-mode states obtained
by taking a fixed number, $M$, of infinitely entangled ancillary
bonds (EPR pairs) shared by adjacent sites, and applying an
arbitrary $2M \rightarrow 1$ Gaussian operation on each site
$i=1,\ldots,N$. Such a construction, more properly definable as a
``valence bond'' picture for Gaussian states, can be better
understood by resorting to the Jamiolkowski isomorphism between
quantum operations and quantum states \cite{giedke}. In this
framework, one starts with a chain of $N$ Gaussian states of $2M+1$
modes (the {\em building blocks}). The global Gaussian state of the
chain is described by a CM $\gr\Gamma = \bigoplus_{i=1}^N
\gr\gamma^{[i]}$. As the interest in GVBS lies mainly in their
connections with ground states of Hamiltonians invariant under
translation \cite{GMPS}, we can focus on pure ($\det \gr\gamma^{[i]}
= 1$), translationally invariant ($\gr\gamma^{[i]} \equiv \gr\gamma
\, \forall i$) GVBS. Moreover, in this work we consider
single-bonded GVBS, i.e.~with $M=1$. This is also physically
motivated in view of experimental implementations of GVBS, as more
than one EPR bond would result in a building block with five or more
correlated modes, which appears technologically demanding.

\begin{figure}[t!]
\centering{
\includegraphics[width=8cm]{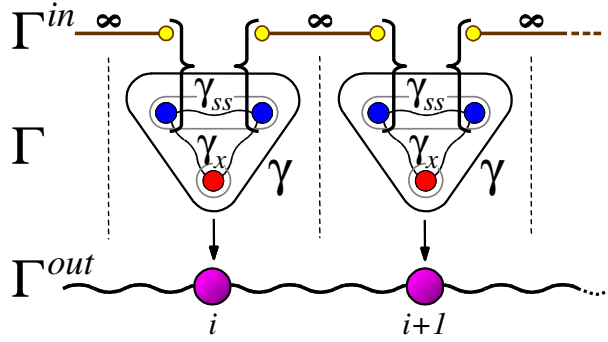} \caption{Gaussian valence bond states.
$\gr\Gamma^{in}$ is the state of $N$ EPR bonds and $\gam$ is the
three-mode building block. After the EPR measurements (depicted as
curly brackets), the chain of modes $\gr\gamma_x$ collapses into a
Gaussian valence bond state with global state $\gr\Gamma^{out}$. See
also Ref.~\cite{marie}.}} \label{fiocco}
\end{figure}

Under the considered  prescriptions, the building block $\gr\gamma$
is a pure Gaussian state of three modes. As we aim to construct a
translationally invariant state, it is convenient to consider a
$\gr\gamma$ whose first two modes, which will be combined with two
identical halves of consecutive EPR bonds (see Fig.~\ref{fiocco}),
have the same reduced CM. This yields a pure, three-mode Gaussian
building block with the property of being {\em bisymmetric}
\cite{adescaling}, that is with a CM invariant under permutation of
the first two modes. This choice of the building block is further
justified by the fact that, among all pure three-mode Gaussian
states, bisymmetric states maximize the genuine tripartite
entanglement \cite{3modi}: no entanglement is thus wasted in the
projection process. The $6 \times 6$ CM $\gr\gamma$ of the building
block can be written as follows in terms of $2\times2$ submatrices,
\begin{equation}\label{bblock}
\gr\gamma = \left(
           \begin{array}{ccc}
             \gam_{s} & \eps_{ss} & \eps_{sx} \\
             \eps_{ss}^T & \gam_{s} & \eps_{sx} \\
             \eps_{sx}^T & \eps_{sx}^T & \gam_{x} \\
           \end{array}
         \right)\!.
\end{equation}
The $4\times 4$ CM of the first two modes (each of them having
reduced CM $\gr\gamma_s$) will be denoted by $\gr\gamma_{ss}$, and
will be regarded as the {\em input} port of the building block. On
the other hand, the CM $\gr\gamma_x$ of mode $3$ will play the role
of the {\em output} port. The intermodal correlations are encoded in
the off-diagonal $\eps$ matrices. Without loss of generality, we can
assume $\gam$ to be, up to local unitary operations, in the standard
form \cite{3modi} with
\begin{eqnarray}
  &&\gam_s = {\rm diag}\{s,\,s\}\,,\quad \gam_x = {\rm diag}\{x,\,x\}\,, \label{bsform}\\
    &&\eps_{ss} = {\rm diag}\{t_+,\,t_-\}\,,\quad \eps_{sx} = {\rm diag}\{u_+,\,u_-\}\,; \nonumber \\
 && t_{\pm}=\frac{1}{4s}\left[x^2-1 \pm \sqrt{16 s^4 - 8 (x^2 + 1) s^2 + (x^2 -
1)^2}\right]\,, \nonumber \\
  &&u_\pm = \frac{1}{4} \sqrt{\frac{x^2 - 1}{s x}} \left[\sqrt{(x - 2
s)^2 - 1} \pm \sqrt{(x + 2 s)^2 - 1}\right]\,. \nonumber
\end{eqnarray}

The valence bond construction works as follows (see
Fig.~\ref{fiocco}). The global CM $\gr\Gamma = \bigoplus_{i=1}^N
\gr\gamma$ acts as the projector from the  state $\gr\Gamma^{in}$ of
the $N$ ancillary EPR pairs, to the final $N$-mode GVBS
$\gr\Gamma^{out}$. This is realized by collapsing the  state
$\gr\Gamma^{in}$, transposed in phase space, with the `input port'
$\gr\Gamma_{ss}=\bigoplus_i \gr\gamma_{ss}$ of $\gr\Gamma$, so that
the `output port' $\gr\Gamma_{x} = \bigoplus_i \gr\gamma_x$ turns
into the desired $\gr\Gamma^{out}$. Here collapsing means that, at
each site, the two two-mode states, each constituted by one mode
($1$ or $2$)  of $\gr\gamma_{ss}$ and one half of the EPR bond
between site $i$ and its neighbor ($i-1$ or $i+1$, respectively),
undergo an ``EPR measurement'' i.e.~are projected onto the
infinitely entangled EPR state \cite{giedke,GMPS}. An EPR pair
between modes $i$ and $j$ can be described, see \eq{tmss}, as a
two-mode squeezed state $\sig_{i,j}(r)$ in the limit of infinite
squeezing ($r \rightarrow \infty$). The input state is then
$\gr\Gamma^{in} = \lim_{r \rightarrow \infty} \bigoplus_{i}^{N}
\gr\sigma_{i,i+1}(r)$, where we have set periodic boundary
conditions so that $N+1 = 1$ in labeling the sites. The projection
corresponds mathematically to taking a Schur complement (see
Refs.~\cite{marie,GMPS,giedke} for details), yielding an output pure
GVBS of $N$ modes on a ring with a CM
\begin{equation}\label{cmout}
\gr\Gamma^{out} = \gr\Gamma_{x} - \gr\Gamma_{sx}^T (\gr\Gamma_{ss} +
\gr\theta \gr\Gamma^{in} \gr\theta)^{-1} \gr\Gamma_{sx}\,,
\end{equation}
where $\gr\Gamma_{sx}=\bigoplus_i \gr\gamma_{sx}$, and $\gr\theta
=\bigoplus_i {\rm diag}\{1,\,-1,\,1,\,-1\}$ represents transposition
in phase space ($\hat q_i \rightarrow \hat q_i,\,\hat p_i
\rightarrow - \hat p_i$).

\begin{figure}[t!]
\centering{
\includegraphics[width=12cm]{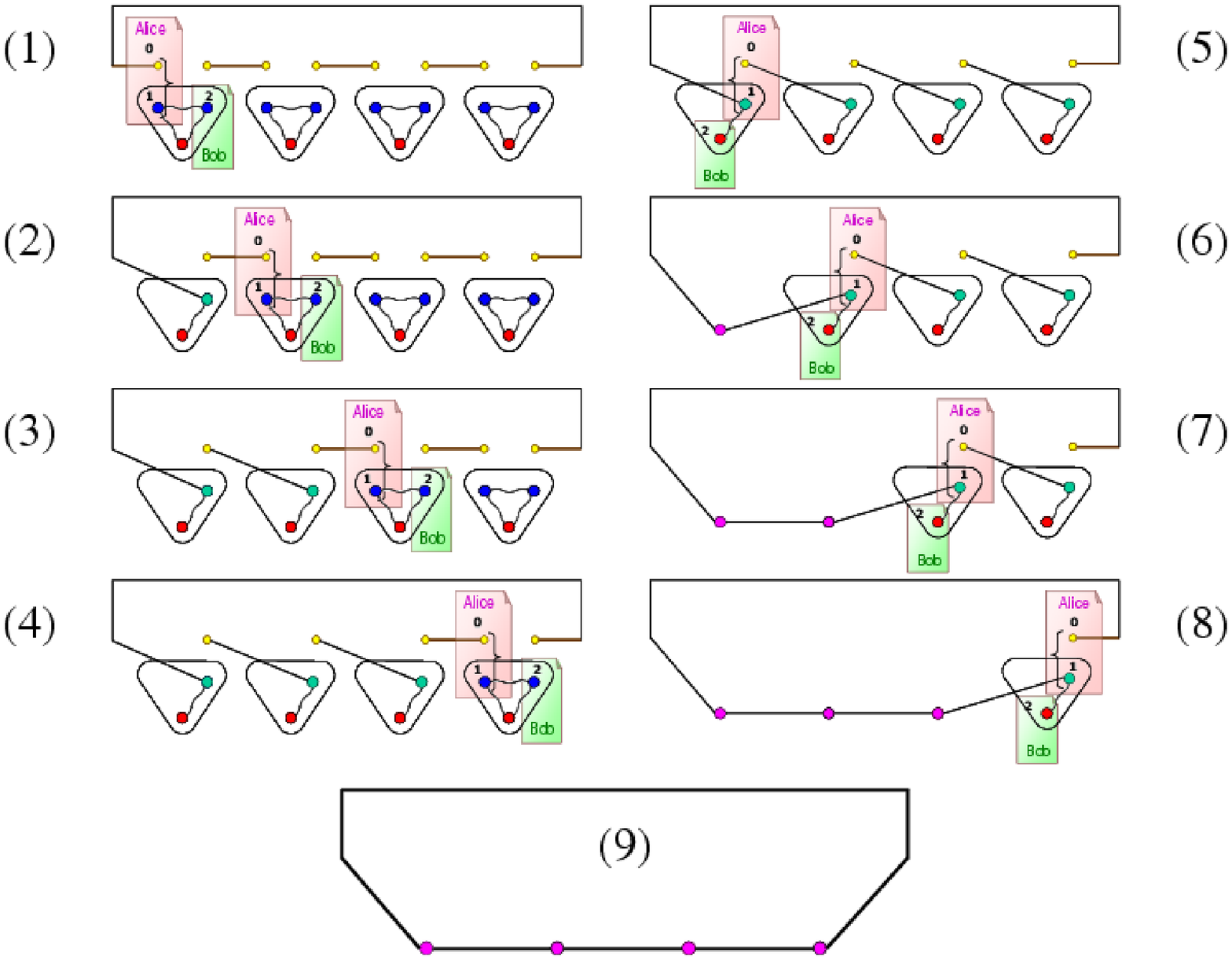} \caption{How a Gaussian valence bond state is created via continuous-variable
entanglement swapping. At each step, Alice attempts to teleport her
mode 0 (half of an EPR bond, depicted in yellow) to Bob, exploiting
as an entangled resource two of the three modes of the building
block (denoted at each step by 1 and 2). The curly bracket denotes
homodyne detection, which together with classical communication and
conditional displacement at Bob's side achieves teleportation. The
state will be approximately recovered in mode 2, owned by Bob. Since
mode 0, at each step, is entangled with the respective half of an
EPR bond, the process swaps entanglement from the ancillary chain of
the EPR bonds to the modes in the building block. The picture has to
be followed column-wise. For ease of clarity, we depict the process
as constituted by two sequences: in the first sequence [frames (1)
to (4)] modes 1 and 2 are the two input modes of the building block
(depicted in blue); in the second sequence [frames (5) to (8)] modes
1 and 2 are respectively an input and an output mode of the building
block. As a result of the multiple entanglement swapping [frame (9)]
the chain of the output modes (depicted in red), initially in a
product state, is transformed into a translationally invariant
Gaussian valence bond state, possessing in general multipartite
entanglement among all the modes (depicted in magenta).}}
\label{vbswap}
\end{figure}

Within the building block picture, the valence bond construction can
be {\em in toto} understood as a multiple CV entanglement swapping
\cite{entswap}, as shown in Fig.~\ref{vbswap}: the GVBS is created
as the entanglement in the bonds is swapped to the chain of output
modes via teleportation \cite{tele} through the input port of the
building blocks. It is thus clear that at a given initialization of
the output port (i.e.~at fixed $x$), changing the properties of the
input port (i.e.~varying $s$), which corresponds to implementing
different Gaussian projections from the ancillary space to the
physical one, will affect the structure and entanglement properties
of the target GVBS. This link is explored in the following section.

\section{Entanglement distribution} \label{secdist}
\begin{figure}[t!]
\centering{ \includegraphics[width=13cm]{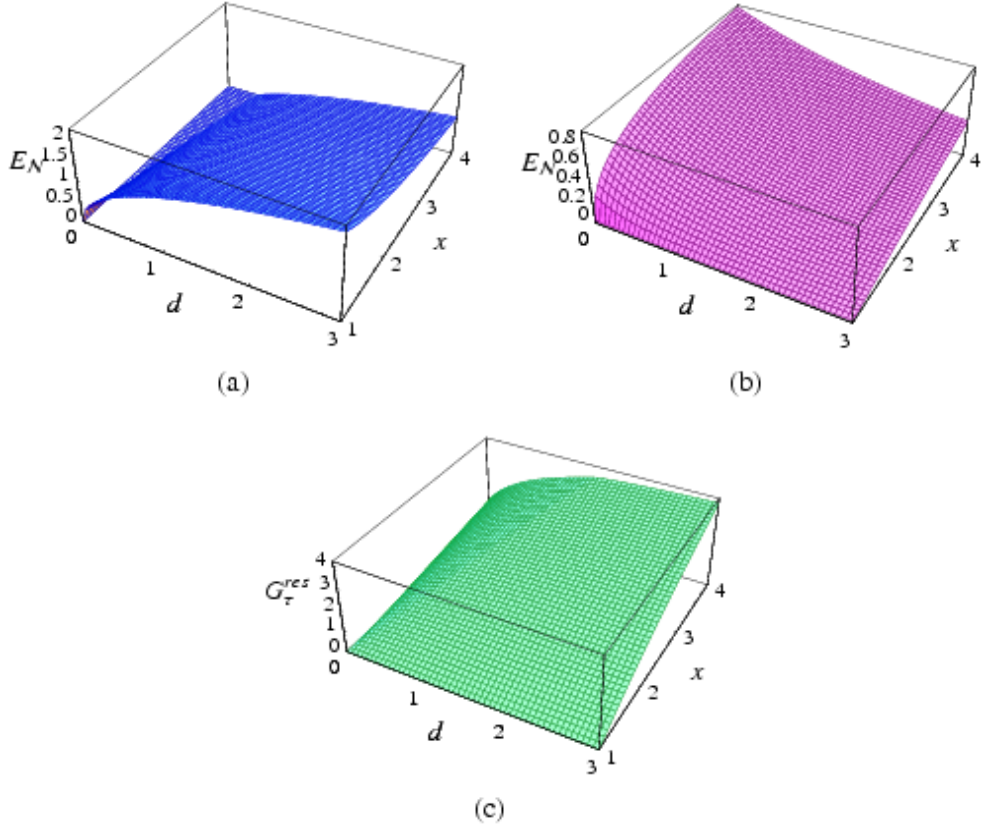}
\caption{Entanglement properties of the three-mode building block
$\gr\gamma$, \eq{bblock}, of the Gaussian valence bond construction,
as functions of the standard form covariances $x$ and $d \equiv
s-s_{\min}$. {\rm (a)} Bipartite entanglement, as quantified by the
logarithmic negativity, between the first two input-port modes 1 and
2; {\rm (b)} Bipartite entanglement, as quantified by the
logarithmic negativity, between each of the first two modes and the
output-port mode 3; {\rm (c)} Genuine tripartite entanglement, as
quantified by the residual Gaussian contangle
\cite{contangle,3modi}, among all the three modes.}}
 \label{fiblocco}
\end{figure}

In Ref.~\cite{marie} the quantum correlations of GVBS of the form
\eq{cmout} have been studied, and related to the entanglement
properties of the building block $\gr\gamma$. Let us first recall
the characterization of entanglement in the latter. As a consequence
of the uncertainty principle \eq{bonfide}, the CM \eq{bblock} of the
building block describes a physical state if \cite{3modi}
\begin{equation}\label{xs}
x \ge 1\,,\quad s\ge s_{\min}\equiv\frac{x+1}{2}\,.
\end{equation}
Let us keep the output parameter $x$ fixed. Straightforward
applications of the PPT separability conditions, and consequent
calculations of the logarithmic negativity \eq{logneg}, reveal that
the entanglement in the CM $\gr\gamma_{ss}$ of the first two modes
(input port) is monotonically increasing as a function of $s$,
ranging from the case $s=s_{\min}$ when $\gr\gamma_{ss}$ is
separable to the limit $s \rightarrow \infty$ when the block
$\gr\gamma_{ss}$ is infinitely entangled. Accordingly, the
entanglement between each of the first two modes $\gam_s$ of
$\gr\gamma$ and the third one $\gam_x$ decreases with $s$. One can
also show that the genuine tripartite entanglement in the building
block increases with the difference $s-s_{\min}$ \cite{3modi}. The
entanglement properties of the building block are summarized in
Fig.~\ref{fiblocco}.

\begin{figure}[t!]
\centering{
\includegraphics[width=9cm]{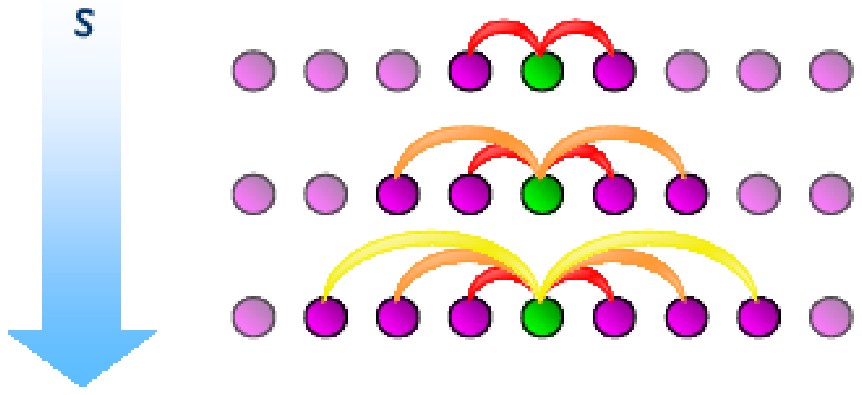}
\caption{Pictorial representation of the entanglement between a
probe (green) mode and its neighbor (magenta) modes on an harmonic
ring with an underlying valence bond structure. As soon as the
parameter $s$ (encoding entanglement in the input port of the
valence bond building block) is increased, pairwise entanglement
between the probe mode and its farther and farther neighbors
gradually appears in the corresponding output Gaussian valence bond
states. By translational invariance, each mode exhibits the same
entanglement structure with its respective neighbors. In the limit
$s \rightarrow \infty$, every single mode becomes equally entangled
with every other single mode on the ring, independently of their
relative distance: the Gaussian valence bond state is in this case
fully symmetric.}} \label{figtalk}
\end{figure}

The main question addressed in Ref.~\cite{marie} is how the initial
entanglement in the building block $\gr\gamma$ redistributes in the
Gaussian MPS $\gr\Gamma^{out}$. The answer is that the more
entanglement one prepares in the input port $\gr\gamma_{ss}$, the
longer the range of pairwise quantum correlations in the output GVBS
is, as pictorially shown in Fig.~\ref{figtalk}.

In more detail, let us consider first a building block  $\gr\gamma$
with $s=s_{\min}=(x+1)/2$. In this case,  a separability analysis
shows that, for an arbitrary number $N$ of modes in the GVBS chain,
the target state $\gr\Gamma^{out}$ exhibits bipartite entanglement
only between nearest neighbor modes, for any value of $x>1$ (for
$x=1$ we trivially obtain a product state). In fact, each reduced
two-mode block $\gr\gamma^{out}_{i,j}$ is separable for $|i-j|>1$.

With increasing $s$ in the choice of the building block, one finds
that in the target GVBS the correlations start to extend smoothly to
distant modes. A series of thresholds $s_k$ can be found such that
for $s > s_k$, two given modes $i$ and $j$ with $|i-j| \le k$ are
entangled. While trivially $s_1(x) = s_{\min}$ for any $N$ (notice
that nearest neighbors are entangled also for $s=s_1$), the
entanglement boundaries for $k>1$ are in general different functions
of $x$, depending on the number of modes. We observe however a
 certain regularity in the process:  $s_k(x,N)$ always increases with the
 integer $k$.  Very remarkably, this means that the maximum range of bipartite
entanglement  between two modes, or equivalently the maximum
distribution of multipartite entanglement, in a GVBS on a
translationally invariant ring, is {\em monotonically} related to
the amount of entanglement in the reduced two-mode input port of the
building block \cite{marie}. Moreover, no complete transfer of
entanglement to more distant modes occurs:  closer sites remain
still entangled even when correlations between farther pairs arise.

The most interesting feature is perhaps obtained when infinite
entanglement is fed in the input port ($s \rightarrow \infty$): in
this limit, the output GVBS turns out to be a fully symmetric,
permutation-invariant, $N$-mode Gaussian state. This means that each
individual mode is {\em equally entangled} with any other, no matter
how distant they are \cite{marie}. These states, being thus built by
a symmetric distribution of infinite pairwise entanglement among
multiple modes, achieve maximum genuine multiparty entanglement
among all Gaussian states (at a given energy) while keeping the
strongest possible bipartite one in any pair, a property known as
monogamous but promiscuous entanglement sharing \cite{contangle}.

Keeping Fig.~\ref{vbswap} in mind, we can conclude that having the
two input modes initially entangled in the building blocks,
increases the efficiency of the entanglement-swapping mechanism,
inducing  correlations between distant modes on the GVBS chain,
which enable to store and distribute joint information. In the
asymptotic limit of an infinitely entangled input port of the
building block, the entanglement range in the target GVBS states is
engineered to be maximum, and communication between any two modes,
independently of their distance, is enabled nonclassically.
 In the next sections, we
investigate the possibility of producing GVBS with linear optics,
and discuss with a specific example the usefulness of such resource
states for multiparty CV quantum communication protocols such as
telecloning \cite{teleclon} and teleportation networks
\cite{network}.

\section{Optical implementation of Gaussian valence bond states}
\label{secopt}

The power of describing the production of GVBS in terms of physical
states, the building blocks, rather than in terms of arbitrary
non-unitary Gaussian maps, lies not only in the immediacy of the
analytical treatment. From a practical point of view, the recipe of
Fig.~\ref{fiocco} can be directly implemented to produce GVBS
experimentally in the domain of quantum optics. We first note that
the EPR measurements are realized by the standard toolbox of a
beamsplitter plus homodyne detection \cite{giedke}, as demonstrated
in several CV teleportation experiments \cite{telexp}.

\begin{figure}[t!]
\centering{\includegraphics[width=8cm]{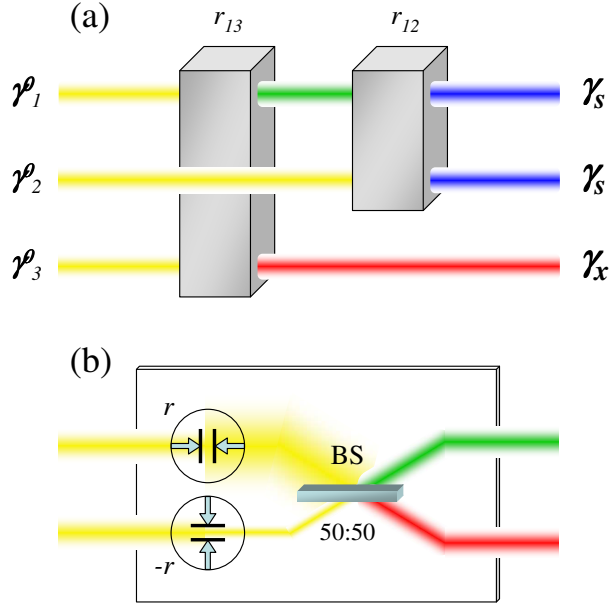} \caption{Optical
production of bisymmetric three-mode Gaussian states, used as
building blocks for the valence bond construction. (a) Three initial
vacuum modes are entangled through two sequential twin-beam boxes,
the first (parametrized by a squeezing degree $r_{13}$) acting on
modes $1$ and $3$, and the second (parametrized by a squeezing
degree $r_{12}$) acting on the transformed mode $1$ and mode $2$.
The output is a pure three-mode Gaussian state whose covariance
matrix is equivalent, up to local unitary operations, to the
standard form given in \eq{bblock}. (b) Detail of the entangling
twin-beam transformation. One input mode is squeezed in a
quadrature, say momentum,  of a degree $r$ (this transformation is
denoted by stretching arrows
$\rightarrow\!\!\!|\,|\!\!\!\leftarrow$); the other input mode is
squeezed in the orthogonal quadrature, say position, of the same
amount (this anti-squeezing transformation is denoted by the
corresponding rotated symbol). Then the two squeezed modes are
combined at a 50:50 beam-splitter. If the input modes are both in
the vacuum state, the output is a pure two-mode squeezed Gaussian
state, with entanglement proportional to the degree of squeezing
$r$.}} \label{bbopt}
\end{figure}

The next ingredient to produce a $N$-mode GVBS is constituted by $N$
copies of the three-mode building block $\gam$. We provide here an
easy scheme (see also Refs.~\cite{teleclon,mygeneric}) to realize
bisymmetric three-mode Gaussian states of the form \eq{bblock}. As
shown in Fig.~\ref{bbopt}(a), one can start from three vacuum modes
and first apply a twin-beam operation to modes $1$ and $3$,
characterized by a squeezing $r_{13}$, then apply another twin-beam
operation to modes $1$ and $2$, parametrized by $r_{12}$. The
symplectic operation describing the twin-beam transformation
(two-mode squeezing plus balanced beam splitter) is given by
\eq{twin} and pictorially represented in Fig.~\ref{bbopt}(b). The
output of this optical network is a pure, bisymmetric, three-mode
Gaussian state with a CM $\gam_B =
T_{12}(r_{12})T_{13}(r_{13})T_{13}^T(r_{13})T_{12}^T(r_{12})$ of the
form \eq{bblock}, with
\begin{eqnarray}
  &&\gam_s = {\rm diag}\left\{\frac{1}{2} e^{-2 r_{12}} \left(e^{4 r_{12}} \cosh \left(2 r_{13}
  \right) + 1\right),\, \frac{1}{2} e^{-2 r_{12}} \left(\cosh \left(2
r_{13}\right) \ + e^{4 r_{12}}\right)\right\}\,, \nonumber \\
&& \gam_x = {\rm diag} \left\{\cosh \left(2 r_{13}\right),\, \cosh\left(2 r_{13}\right)\right\}\,, \nonumber \\
    &&\eps_{ss} = {\rm diag} \left\{\frac{1}{2} e^{-2 r_{12}} \left(e^{4 r_{12}} \cosh \left(2 r_{13}
    \right) - 1\right),\, \frac{1}{2} e^{-2 r_{12}} \left(\cosh \left(2
r_{13}\right) \
- e^{4 r_{12}}\right)\right\}\,,   \nonumber \\
&& \eps_{sx} = {\rm diag} \left\{\sqrt{2} e^{r_{12}} \cosh
\left(r_{13}\right) \sinh \left(r_{13}\right),\, -\sqrt{2}
e^{-r_{12}} \cosh \left(r_{13}\right) \sinh
\left(r_{13}\right)\right\}\,.
 \nonumber \\
 &\label{bsformopt}&
\end{eqnarray}
By means of  {\em local} symplectic operations (unitary on the
Hilbert space), like additional single-mode squeezings, the CM
$\gam_B$ can be brought in the standard form of \eq{bsform}, from
which one has $$r_{13}=\arccos\left(\frac{\sqrt{x +
1}}{\sqrt{2}}\right)\,,\quad r_{12}=\arccos\sqrt{\frac{\sqrt{-x^3 +
2 x^2 + 4 s^2 x - x}}{4 x} + \frac{1}{2}}\,.$$ For a given $r_{13}$
(i.e.~at fixed $x$), the quantity $r_{12}$ is a monotonic function
of the standard-form covariance $s$, so this squeezing parameter
which enters in the production of the building block (see
Fig.~\ref{bbopt}) directly regulates the entanglement distribution
in the target GVBS, as discussed in Sec.~\ref{secdist}.

The only unfeasible part of the scheme seems constituted by the
ancillary EPR pairs. But are {\em infinitely} entangled bonds truly
necessary? In Ref.~\cite{marie} the possibility is considered of
using a $\gr\Gamma^{in}$ given by the direct sum of two-mode
squeezed states of \eq{tmss}, but with finite $r$. Repeating the
previous analysis to investigate the entanglement properties of the
resulting GVBS with finitely entangled bonds, it is found that, at
fixed $(x,s)$, the entanglement in the various partitions is
degraded as $r$ decreases, as somehow expected. Crucially, this does
not affect the connection between input entanglement and output
correlation length. Numerical investigations show that, while the
thresholds $s_k$ for the onset of entanglement between distant pairs
are quantitatively modified  --  a bigger  $s$ is required at a
given $x$ to compensate the less entangled bonds -- the overall
structure stays untouched. This ensures that the possibility of
engineering the entanglement structure in GVBS via the properties of
the building block is robust against imperfect resources, definitely
meaning that the presented scheme is feasible.

Alternatively, one could from the beginning observe that the triples
consisting of two projective measurements and one EPR pair can be
replaced by a single projection onto the EPR state, applied at each
site $i$ between the input mode $2$ of the building block and the
consecutive input mode $1$ of the building block of site $i+1$
\cite{GMPS}. The output of all the homodyne measurements will
conditionally realize the target GVBS.

\section{Telecloning with Gaussian valence bond
resources}\label{secclon}

The protocol of CV quantum {\em telecloning} \cite{teleclon} among
$N$ parties is defined as a process in which one of them (Alice)
owns an unknown coherent state, and wants to distribute her state to
all the other $N-1$ remote parties. The telecloning is achieved by a
succession of standard two-party CV teleportations \cite{tele}
between the sender Alice and each of the $N-1$ remote receivers,
exploiting each time the corresponding reduced two-mode state shared
as resource by the selected pair of parties. The $1 \rightarrow 2$
CV telecloning of unknown coherent states has been recently
demonstrated experimentally \cite{teleclonexp}.

The no-cloning theorem \cite{nocloning} yields that the $N-1$ remote
clones can resemble the original input state only to a certain
extent. The {\em fidelity}, which quantifies the success of a
teleportation experiment, is defined as ${\cal F} \equiv
\bra{\psi^{in}} \rho^{out}\ket{\psi^{in}}$, where ``in'' and ``out''
denote the input and the output state. ${\cal F}$ reaches unity only
for a perfect state transfer, $\rho^{out} =
\ket{\psi^{in}}\!\bra{\psi^{in}}$.

Without using entanglement, by purely classical communication, an
average fidelity of ${\cal F}_{cl}=1/2$ is the best that can be
achieved if the alphabet of input states includes all coherent
states with even weight \cite{bfkjmo}.  The sufficient fidelity
criterion states that, if teleportation is performed with ${\cal F}
> {\cal F}_{cl}$, then the two parties exploited an entangled state
\cite{bfkjmo}. The converse is generally false, i.e. some entangled
resources may yield lower-than-classical fidelities.
 In Ref.~\cite{telepoppy} it has
been shown, however, that if the fidelity is optimized over all
possible local unitary operations performed on the shared Gaussian
resource (which preserve entanglement by definition), then it
becomes {\em equivalent}, both qualitatively and quantitatively, to
the entanglement in the resource.

Let us also recall that the fidelity of CV two-user teleportation
\cite{tele} of arbitrary single-mode Gaussian states with CM
$\gam_{in}$ (equal to the identity for coherent states) exploiting
two-mode Gaussian resources with CM $\gam_{ab} = \left(
              \begin{array}{cc}
                \gam_a & \eps_{ab} \\
                \eps_{ab}^{\sf T} & \gam_b \\
              \end{array}
            \right)$, can be computed \cite{fiuratele} as
\begin{equation}\label{ficm}
{\cal F} = \frac2{\sqrt{\det \gr\Sigma}}\,,\qquad \gr\Sigma \equiv
2\, \gam_{in} + \gr{\xi} \gam_a \gr{\xi} + \gam_b + \gr{\xi}
\eps_{ab} + \eps_{ab}^{\sf T} \gr{\xi}\,,
\end{equation}
with $\gr{\xi} = {\rm diag}\{-1\,,1\}$. We can now consider the
general setting of $1\rightarrow N-1$ telecloning, where  $N$
parties share a $N$-mode GVBS as an entangled resource, and one of
them plays the role of Alice (the sender) distributing imperfect
copies of unknown coherent states to all the $N-1$ receivers. For
any $N$, the fidelity can be easily computed from the reduced
two-mode CMs via \eq{ficm} and will depend, for translationally
invariant states, on the relative distance between the two
considered modes.

In this work we focus on a practical example of a GVBS on a
translationally invariant harmonic ring, with $N=6$ modes. As shown
in the previous section, these states can be produced with the
current optical technology. They are completely characterized, up to
local unitary operations, by a $12 \times 12$ CM analytically
obtained from \eq{cmout} by considering the building block in
standard form \eq{bblock}, whose elements are algebraic functions of
$s$ and $x$ here omitted for brevity (as no particular insight is
gained from their explicit expressions). First of all we can
construct the reduced CMs $\gr\gamma^{out}_{i,i+k}$ of two modes
with distance $k$, and evaluate for each $k$ the respective
symplectic eigenvalue $\tilde\nu^{i,i+k}_-$ of the corresponding
partial transpose. The entanglement condition $s > s_k$ will
correspond to the inequality $\tilde\nu^{i,i+k}_- < 1$. With this
conditions one finds that $s_2(x)$ is the only acceptable solution
to the equation: $72 s^8 - 12 (x^2 + 1) s^6 + (-34 x^4 + 28 x^2 -
34) s^4 + (x^6 - 5 x^4 - 5 x^2 + 1) s^2 + (x^2 - 1)^2 (x^4 - 6 x^2 +
1)=0$, while for the next-next-nearest neighbors threshold one has
simply $s_3(x)=x$. This enables us to classify the entanglement
distribution and, more specifically, to observe the interaction
scale in the GVBS $\gr\Gamma^{out}$: as discussed in
Sec.~\ref{secdist} and explicitly shown in Ref.~\cite{marie}, by
increasing initial entanglement in $\gr\gamma_{ss}$ one can
gradually switch on pairwise quantum correlations between more and
more distant sites.

\begin{figure}[t!]
\centering{
\includegraphics[width=13cm]{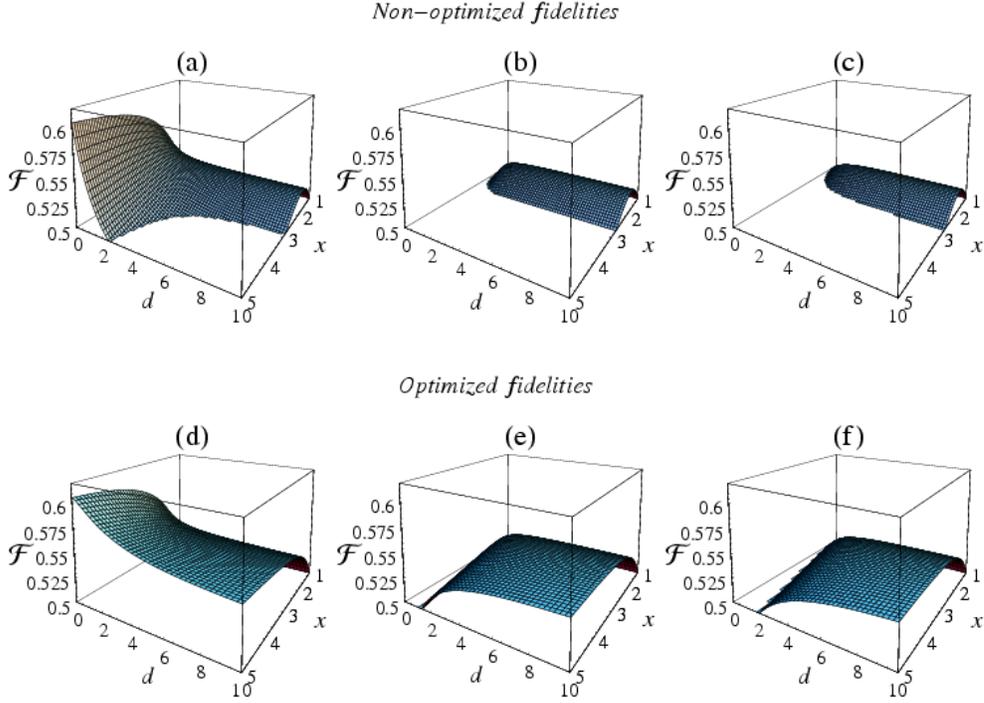}
\caption{$1\rightarrow 5$ quantum telecloning of unknown coherent
states exploiting a six-mode translationally invariant Gaussian
valence bond state as a shared resource. Alice owns mode $i$.
Fidelities $\cal F$ for distributing clones to modes $j$ such as
$k=|i-j|$ are plotted for  $k=1$ [(a),(d)];  $k=2$ [(b),(e)]; and
 $k=3$ [(c),(f)], as functions of the local invariants $s$ and $x$
 of the building block. In the first row [(a)--(c)]
 the fidelities are achieved exploiting the non-optimized
 Gaussian valence bond resource in standard form.
In the second row [(d)--(f)] fidelities optimized over local unitary
operations on the resource are displayed, which are equivalent to
the entanglement in the corresponding reduced two-mode states (see,
as a comparison, Fig.~3 in Ref.~\cite{marie}). Only nonclassical
values of the fidelities (${\cal F} > 0.5$) are shown.}}
\label{televbs}
\end{figure}

Accordingly, it is now interesting to test whether this entanglement
is useful to achieve nonclassical telecloning towards distant
receivers. In this specific instance, Alice will send two identical
(approximate) clones to her nearest neighbors, two other identical
clones (with in principle different fidelity than the previous case)
to her next-nearest neighbors, and one final clone to the most
distant site. The fidelities for the three transmissions can be
computed from \eq{ficm} and are plotted in Fig.~\ref{televbs}(a).
For $s=s_{\min}$, obviously, only the two nearest neighbor clones
can be teleported with nonclassical fidelity, as the reduced states
of more distant pairs are separable. With increasing $s$ also the
state transfer to more distant sites is enabled with nonclassical
efficiency, but not in the whole region of the space of parameters
$s$ and $x$ in which the corresponding two-mode resources are
entangled.

As mentioned before, one can optimize the telecloning fidelity
considering resources prepared in a different way but whose CM can
be brought by local unitary operations (single-mode symplectic
transformations) in the standard form of \eq{cmout}. For GVBS
resources, this local-unitary freedom can be transferred to the
preparation of the building block. A more general $\gam$ locally
equivalent to the standard form given in \eq{bsform}, can be
realized by complementing the presented state engineering scheme for
the three-mode building block as in \eq{bsformopt} [see
Fig.~\ref{bbopt}(a)], with additional single-mode rotations and
squeezing transformations aimed at increasing the output fidelity in
the target GVBS states, while keeping both the entanglement in the
building block and consequently the entanglement in the final GVBS
unchanged by definition.

The optimal telecloning fidelity, obtained in this way exploiting
the results of Ref.~\cite{telepoppy}, is plotted in
Fig.~\ref{televbs}(b) for the three teleportations between modes $i$
and $j$ with $k=|i-j|=1,\,2,\,3$. In this case, one immediately
recovers a non-classical fidelity as soon as the separability
condition $s\le s_k$ is violated in the corresponding resources.
Moreover, the optimal telecloning fidelity at a given $k$ is itself
a quantitative measure of the entanglement in the reduced two-mode
resource, being equal to \cite{telepoppy}
\begin{equation}\label{fiopt2}
{\cal F}^{opt}_k = 1/({1+{\tilde\nu^{i,i+k}_-}})\,,
\end{equation}
where $\tilde\nu^{i,i+k}_-$ is the smallest symplectic eigenvalue of
the partially transposed CM in the corresponding bipartition. The
optimal fidelity is thus completely equivalent to the entanglement
of formation \eq{eof} and to the logarithmic negativity \eq{logneg}.

In the limit $s \rightarrow \infty$, as discussed in
Sec.~\ref{secdist}, the GVBS become fully permutation-invariant for
any $N$. Consequently, the (optimized and non-optimized) telecloning
fidelity for distributing coherent states is equal for any pair of
sender-receiver parties. These resources are thus useful for $1
\rightarrow N-1$ symmetric telecloning. However, due to the monogamy
constraints on distribution of CV entanglement \cite{contangle},
this two-party fidelity will decrease with increasing $N$, vanishing
in the limit $N \rightarrow \infty$ where the resources become
completely separable. In this respect, it is worth pointing out that
the fully symmetric GVBS resources are more useful for teleportation
networks \cite{network,nets}, where $N-2$ parties first perform
local measurements (momentum detections) on their single-mode
portion of the entangled resource to concentrate as much
entanglement as possible onto the two-mode state of Alice and Bob,
who can accomplish non-classical teleportation (after the outcomes
of the $N-2$ measurements are classically communicated to Bob). In
this case, the optimal fidelity of $N$-user teleportation network is
an estimator of {\em multipartite} entanglement in the shared
$N$-mode resource \cite{telepoppy}, which is indeed a GVBS obtained
from an infinitely entangled building block.

\section{Conclusion}
The valence bond picture is a valuable framework to study the
structure of correlations in quantum states of harmonic lattices. In
fact, the motivation for such a formalism is quite different from
the finite-dimensional case, where valence bond/matrix product
states are useful to efficiently approximate ground states of
$N$-body systems -- generally described by a number of parameters
exponential in $N$ -- with polynomial resources \cite{maria}. In
continuous variable systems, the key feature of GVBS lies in the
understanding of their entanglement distribution as governed by the
properties of simpler structures \cite{marie}. This has also
experimental implications giving a robust recipe to engineer
correlations in many-body Gaussian states from feasible operations
on the building blocks. We have provided a simple scheme to produce
bisymmetric three-mode building blocks with linear optics, and
discussed the subsequent implementation of the valence bond
construction. We have also investigated the usefulness of such GVBS
as resources for nonclassical communication, like telecloning of
unknown coherent states to distant receivers on a harmonic ring. It
would be interesting to employ the valence bond picture to describe
quantum computation with continuous-variable  cluster states
\cite{cvcluster}, and to devise efficient protocols for its optical
implementation.

\ack{This work is supported by MIUR (Italy) and by the European
Union through the Integrated Project RESQ (IST-2001-37559), QAP
(IST-3-015848), SCALA (CT-015714), and SECOQC.}

\end{document}